\documentclass[review]{elsarticle}

\journal{Journal of \LaTeX\ Templates}

\usepackage{graphicx}
\usepackage{amsmath}
\usepackage{algorithmic}

\usepackage[ruled,linesnumbered]{algorithm2e}
\usepackage{multirow}

\usepackage{url}

\begin{document}

\begin{frontmatter}

\title{AAE: An Active Auto-Estimator for Improving Graph Storage}


\author[mymainaddress]{Yu Yan}

\author[mymainaddress]{Man Yang}

\author[mymainaddress]{Hongzhi Wang}
\cortext[mycorrespondingauthor]{Corresponding author}

\author[mymainaddress]{Yuzhuo Wang}

\address[mymainaddress]{Harbin Institute of Technology, Harbin, China}

\begin{abstract}

	Nowadays, graph becomes an increasingly popular model in many real applications. The efficiency of graph storage is crucial for these applications. Generally speaking, the tune tasks of graph storage rely on the database administrators (DBAs) to find the best graph storage. However, DBAs make the tune decisions by mainly relying on their experiences and intuition. Due to the limitations of DBAs's experiences, the tunes may have an uncertain performance and conduct worse efficiency. In this paper, we observe that an estimator of graph workload has the potential to guarantee the performance of tune operations. Unfortunately, because of the complex characteristics of graph evaluation task, there exists no mature estimator for graph workload. We formulate the evaluation task of graph workload as a classification task and carefully design the feature engineering process, including graph data features, graph workload features and graph storage features. Considering the complex features of graph and the huge time consumption in graph workload execution, it is difficult for the graph workload estimator to obtain enough training set. So, we propose an active auto-estimator (AAE) for the graph workload evaluation by combining the active learning and deep learning. AAE could achieve good evaluation efficiency with limited training set.
	We test the time efficiency and evaluation accuracy of AAE with two open source graph data, LDBC and Freebase. Experimental results show that our estimator could efficiently complete the graph workload evaluation in milliseconds. 
	\end{abstract}
	
\end{frontmatter}

\textbf{keywords:}{graph storage, workload evaluation,active learning}

\section{Introduction}

Nowadays, graph becomes an increasingly popular model in many real applications~\cite{dual}. For example, in FaceBook, the relationships among users are stored as property graph. In machine cognitive learning, the knowledge of experience is stored as knowledge graph~\cite{know}. Efficient graph storage is the key step for improving the economic income in these applications by accelerating the workload execution. 

Generally speaking, the graph storage tuning tasks mainly rely the database administrators (DBAs) in real applications. However, DBAs complete the tuning by utilizing their experiences and intuition, which may conduct performance degradation. For example, a graph has two properties, age and gender. There exist many queries about gender property. DBAs may establish an index in gender according to the frequency. But the gender property only has two kinds of values in which the index is useless. It is unreliable to tune the graph storage only relying on manual work. Importantly, in real applications, any performance degradation will influence the economic income. So, we need an estimator to help DBAs guarantee the efficiency of the storage tunes.

\begin{table*}[h]
    \caption{The Cost of Labeling}
    \begin{center}
        \begin{tabular}{|c|c|c|c|c|c|c|}
            \hline
            database\& & Create & Read & Update& Delete & Traversal & Workload \\
            graph data & Operation & Operation & Operation & Operation & Operation & Runtime(hours)\\
            \hline
            \multirow{5}*{New4J+LDBC}
            &38\% &15\% &2\%& 13\% &32\% & 11.57\\
            \cline{2-7}
            & 16\% &31\% &3\%& 3\% &47\% & 31.25 \\
            \cline{2-7}
            & 17\% &25\% &17\%& 10\% &31\% &17.36 \\
            \cline{2-7}
            & 9\% &17\% &9\%& 10\% &31\% & 23.15 \\
            \cline{2-7}
            & 20\% &12\% &10\%& 12\% &46\% & 34.72 \\
            \hline
            \multirow{5}*{New4J+Freebase}
            &20\% &18\% &13\%& 11\% &28\% & 18.52\\
            \cline{2-7}
            & 14\% &12\% &28\%& 11\% &35\% &11.53 \\
            \cline{2-7}
            & 10\%& 49\% &6\%& 11\% &24\% &28.33 \\
            \cline{2-7}
            & 12\% &23\% &4\%& 33\% &28\% &57.87 \\
            \cline{2-7}
            & 19\% &19\% &3\%& 11\% &57\% &173.61 \\
            \hline
        \end{tabular}
    \label{sample}
    \end{center}
    \end{table*}

Unfortunately, existing workload estimators~\cite{sample}~\cite{sample2}~\cite{his}~\cite{machine}~\cite{machine2} which are mainly designed for the sample data models like relational model, key value model and document model is unsuitable for the graph model. For example, the learned model~\cite{machine2} proposed by R.J et al could only process some sample filters, like `range', `in' and etc. while do not support the traversal filters of graph data.
Specifically, due to the complexity of graph workload evaluation, there exists no mature work for this task. 
In this paper, we pay attention to construct an estimator for evaluating the graph workload to guarantee the tune performance in real applications.

Firstly, we conclude three aspect reasons why evaluating graph workload is difficult as follows:

\begin{itemize}
    \item \textbf{Complex Structures} In order to save various entities and relationships, graph data has very complex structures. For example, there exist more than 1000 million edges in LinkedGeoData. And DBpedia has more than 3000 million triples. Big graph has more complex structural features and statistical features~\cite{big}. Evaluating the graph workload in such big graph is very difficult.
    \item \textbf{Diverse Query Operations} To make full use of graph data information, the query patterns in graph are usually diverse including 'create nodes', 'read nodes', 'update nodes', 'delete nodes' and 'traverse nodes'. And every kind has diverse basic operations~\cite{diverse}. Only the 'read nodes' contains four basic operations seen in Table~\ref{table2}. Evaluate graph workload under diverse query operations is very difficult.
    \item \textbf{Dynamic Workloads} In real applications, the workload is changing with the user's requests. The dynamic workloads also bring big challenges to the evaluation method. On the one hand, changing workloads require fast evaluation to guarantee the efficient self-tuning. On the other hand, too many factors in workloads, which have great impact on graph workload execution, need to carefully evaluate. It is necessary for the estimator to quickly complete the evaluation task with high accuracy.
\end{itemize}

Thus, it is necessary to design a special evaluation model for graph workloads. Inspired by the recent machine learning methods~\cite{meet}~\cite{card} in database committee, we firstly observe how to construct an automatic graph workload estimator to carry the above difficulties. We formulate the graph workload evaluation task into a classification task and explore how to utilize deep learning classifiers to complete this evaluation task. The input consists of the graph workload, the graph data, old-graph storage and new-graph storage. The output presents if the new revision of storage structure is better than the old storage structure.

However, deep learning based models need a large number of labeled data. The result of our preliminary investigation in Table~\ref{sample} shows that it is incredible to obtain enough labeled data in limited time. We can find that the first item in Table~\ref{sample} would take 11.57 hours in New4J and LDBC dataset. In order to reduce the cost of data labeling, we combine the active learning structure~\cite{activel} to our estimator and propose an active auto-estimator (AAE) for graph workload to decrease the demand of training sets. Our experimental results prove that only using a bit labeled data can still make good performance. For example, AAE reach 72\% accuracy only with 42\% training set. And our estimator could complete the evaluation task with milliseconds.

The contributions of this paper are summarized as follows:

\begin{itemize}
\item We firstly formulate the definition of graph workload evaluation problem which could help to improve the graph storage and increase the economic efficiency of applications.

\item We carefully design the features of graph data, graph workload and graph storage, containing the structural features and statistical features of graph data, various graph query pattern features, basic storage schema and indexes features.

\item We design a graph workload estimator, AAE which combines the deep learning and active learning. Our model could process the complex features, including data features, workload features and graph storage features. And AAE could quickly complete the graph workload evaluation with the limited training set.

\item We examine our active estimator with  two widely used dataset, LDBC and Freebase in microbenchmark~\cite{ref14}. The experimental result shows that AAE could largely reduce the demand of training dataset. And our estimator could complete evaluation with milliseconds.
\end{itemize}

The structure of this paper is as follows. Section~\ref{related} presents some related works about graph storage and graph query acceleration. Section~\ref{problem} introduces the overview of AAE. Section~\ref{feature} discuss the feature engineering of three key components of AAE, including graph data, graph workload and graph storage. And Section~\ref{active} carefully introduce the active auto-estimator. Finally, we clarify the experimental result in section~\ref{result}, containing experimental dataset and result analysis.

\section{related work}\label{related}
Because of the lacks of the graph workload evaluation methods, we only introduce some related works about graph storage and graph workload acceleration.

\textbf{Graph Storage}
The storage of graph data is divided into native graph database storage~\cite{ref5} and non-native graph database storage~\cite{ref6,ref7}. The native graph database implements point-to-point data structure and indexing; the non-native graph database uses other database systems to store graph data and implement query interfaces. Native graph database systems such as Neo4j, InfiniteGraph, Sparksee~\cite{ref3}. Non-native graph database systems such as OrientDB, ArangoDB and various relational databases. Compared with native storage, non-native storage systems~\cite{ref8,ref9} have more mature  support in terms of concurrency~\cite{ref4}, locks, security, and query optimization.

\textbf{Graph workload Acceleration}
Recently, researchers mainly optimal the graph query by two key aspects. One is to construct indexes. Compared to other data models, graph data has more complex data structure and diverse query operations. For quickly scan the complex graph, researches index the subgraph~\cite{sub}, path~\cite{path} and etc. Different from the sample data model, using the general indexes (like B+ tree, Hash, LSM tree) do not quickly find some complex patterns~\cite{pattern}. Expect the index there also exists some other methods to accelerate graph query. People propose some cache and prefetching methods~\cite{cache}~\cite{simi}~\cite{simi2}~\cite{simi3}~\cite{sime4} based on similarity counting. And recent works~\cite{simi3} prefetch graph query by computing the graph edit distance between queries. Totally, the graph database management urgently need the evaluation modules to accelerate the graph query.

\section{Overview}\label{problem}

Traditional estimators could be divided into two main class, single query based~\cite{machine2} and workload based~\cite{work}.  Considering the demand of real-world settings like FaceBook which could receive millions of queries at one time, we take the total graph workload as the evaluation target. There are two significant influential factors to graph workload performance, graph data and graph storage. In this paper, we formulate these key components of our estimator as follows: 

 \textbf{Graph Data:} $\mathcal{G}$ $=(N,R,P,T)$, where $N$ is the collection of nodes, $R$ is the collection of edges, $P$ is the collection of properties, and $T$ is the property edges.

 \textbf{Graph Workload:} $\mathcal{W}$ $ = \{q_1, q_2, ... q_n\}$, where $q_i$ is a graph query.

 \textbf{Graph Storage:} $\mathcal{S} $ $ = <S_S, S_I>$,where $S_S$ is the basic storage schema and the $S_I$ is the indexes in $S_S$.

 The total cost of graph workload is denoted as follows. $w_i$ is the weight of $q_i$ and $c_i$ is the cost of $q_i$.

 \begin{equation}
    \mathcal{C}ost(G, W, S)  = \sum_{i=1}^n{w_i*c_i}
 \end{equation}

Comparing the workload execution cost of two graph storage plans could help DBAs to stably tune the graph storage. In order to improve graph storage, we formulate the graph workload evaluation task as a classification task.

 \textsf{Definition}(the graph workload evaluation task):
 Give the graph data $G$ and workload $W$,
We define the cost a graph workload evaluation task as follows:
\begin{equation}
    \label{eq6}
    AAE(G, W, S_old, S_new)=\left\{
    \begin{aligned}
    1 & , & \mathcal{C}ost(G, W, S_old) > \\
     &  & \mathcal{C}ost(G, W, S_new),  \\
    0 & , & \mathcal{C}ost(G, W, S_old) \leq \\
    &  & \mathcal{C}ost(G, W, S_new).
    \end{aligned}
    \right.
    \end{equation}

 The input of AAE consists of the graph workload, the graph data, the old graph storage and the new graph storage. The output presents if the new graph storage is better than the old one in the current graph data and graph workload. Our AAE could help DBAs to improve the graph storage by quickly evaluating graph workloads. 

\begin{figure}[htbp]
    \centering
    \includegraphics[width=3in]{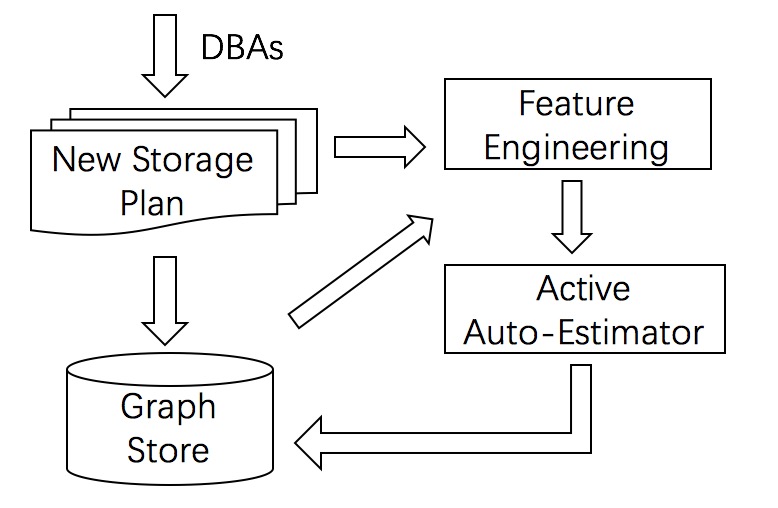}
    \caption{The Workflow of AEE}
\label{overview}
\end{figure}

 As figure~\ref{overview} shows, after receiving the graph storage tune from DBAs, our system implements the evaluation to judge if the tune is positive. If so, this tune from DBAs could be implemented to database. Our estimator could help DBAs to prevent the performance degradation.

\section{Feature Engineering}\label{feature}
Feature engineering is the key step of constructing the graph workload estimator. 
Comparing with the evaluation for other sample data models (like relation, key-value), the graph evaluation has more complex characteristics, such as the complex graph structure, diverse query operations and dynamic workloads. In this section, we introduce the feature engineering of constructing graph workload estimator, including the graph data, the graph workload and the graph storage.

\textbf{Dataset Feature Extraction:} Generally Speaking, the characteristics of graph data mainly include statistical characteristics and structural characteristics. In this paper, we pay more attention to these two kinds of features for the representation of graph data. Considering the scale and complexity of the graph data, we conclude that some vital statistical features, including data size, number of nodes, number of edges.  And some key structural features include the number of nodes or edges' type, the number of properties and the property values of node or edge. Our estimator which use both statistical features and structural features has more complete representation for graph dataset.

\textbf{Workload Feature Extraction:} 
Graph has various basic query operations. We carefully divide these operations into five types as listed in Table~\ref{table2}, containing create, read, update, delete and traverse. We can find that different from other sample data models (like relation, key-value), graph has more plentiful operations. We conclude these operations by totally analyzing the graph query and existing works~\cite{cache,diverse}. And the rate of these operations in workload has vital influence for graph workload evaluation. For example, in native graph storage, traversal graph queries usually have higher efficiency than read queries. While in relational graph storage, read queries usually have higher efficiency than traversal queries~\cite{dual}. By extracting the characteristics of the workload, we can identify the main types of operations, and improve the accuracy of our estimator. Through these workload information, we can have a more detailed judgement of the graph storage. Finally, we utilize the rate of each basic query operation and the frequency of properties as the representation of the graph workloads.

\begin{table}[]
    \caption{graph query operations}
    \begin{center}
    \begin{tabular}{|p{0.5in}|p{1.5in}|p{3in}|}
    \hline
    \textbf{Type}&\textbf{Query Function}&\textbf{Description}\\
    \hline
    &addVertex(G, n)&Create a new node n to graph G\\
    Create&addEdge(G, e)&Create an edge e from n1 to n2 in Graph G\\

    & addProperty(G.N.n,p)&Create a property to nodes or edges\\

    \hline
    &getCount(G.N)&Count the total number of nodes or edges\\

    &getProperty(G.E)&Get all edge or nodes properties in graph G\\
    Read& findProperty(G.N, p)&search nodes or edges in G.N according to the property p\\

    &find(G.N, n)&Search nodes or edges in G.N according to the given condition \\

    \hline
    Update&setProperty(G.N,n,p)&Update the properties of nodes or edges\\

    \hline
    &removeVertex(G, n)&Delete the node according to the identifier n\\
    &removeEdge(G, e)&Delete the edge according to the identifier e\\
    Delete&removeProperty(G.N,n,p)&Delete the property p of nodes or edges\\

    \hline
    &in(G.N, n)&Query all nodes adjacent to the node n with an incoming edge\\
    &out(G.N, n)&Query all nodes adjacent to the node n with an outcoming edge\\
    &all(G.N, n)&Query all nodes adjacent to the node n\\
    &TFilter(count(in(G.N,n)) $\geq$ k) &Filter all the nodes which have at least k-incoming-degree\\

    Traverse

    &allinPath(BFS(G,n))&Search all the nodes reached via breadth-First traversal from node n\\
    &allinPath(BFS(G,n,l))&Same as above,but with edges' labels limitation\\
    &shortPath(G,n1,n2)&Get the shortest path from node n1 to node n2 without weight\\
    &shortPath(G,n1,n2,l)&Same as above,but with edges' labels limitation \\
    \hline
    \end{tabular}
    \label{table2}
    \end{center}
    \end{table}

\textbf{Storage Solution Features:} In order to query graph data quickly, there are some different kinds of graph store engines, including New4J, Titan, ArangoDB and etc. Also, there are more other features in corresponding engines, such as index support, query optimizer and other feature of distribution design. In this paper, we choose the basic stores and indexes as the total represented features for graph storage solution. Because of the basic stores and indexes information are not numeral features, we utilize the one-hot method~\cite{end} to encode them. And one-hot method could largely retain the original characteristics. Specifically, we encode every kind of engines as a one-hot vector. For index information, we consider the indexes on graph node properties and edge properties. We also utilize the one-hot vector to identify if a property has the index.

\section{Active Auto-Estimator}\label{active}
It is difficult to obtaining a large mount of training set in graph workload evaluation task with limited time. We propose an active-based solution to reduce the time consumption of training model. In this section, we introduce the active auto-estimator module.

\subsection{Active Structure}

Generally speaking, the classifiers based deep learning need a large number of labeled dataset for model training. However, as our preliminary results in Table~\ref{sample}, graph workload execution requires a lot of time and calculation resources, it is incredible for obtaining enough labeled dataset by implement graph queries in graph database with certain storage and graph. In order to reduce the cost of obtaining labeled data, we design an active based model to evaluate graph storage. The active learning~\cite{ref18} means training models only with partial dataset. Firstly, we sample some data from the unlabeled data. And next, we implement data training with these sample data. Then, we use the trained model to evaluate unlabeled data. If the accuracy reach the threshold, we stop this interaction. And if the accuracy is not enough, we would sample the poor data point which does not perform well under the trained model to retrain the classifier. Repeat the above process, we could obtain an evaluation model with smaller training set finally.

\begin{algorithm}[t]
	\caption{Active Learning}\label{algor}
	\SetAlgoLined
    \SetKwInOut{Input}{input}
    \Input{
        \ $U$ is the unlabeled dataset
        \\
        \ $L$ is the labeled dataset
        \\
        \ $\Theta$ is the parameters of deep classifiers
        \\
        \ $T$ is the threshold of the estimation
    }
    \vspace{1mm}

    \SetKwInOut{Output}{output}
    \Output{
        \ $\Theta$ the trained parameters

    }
	\vspace{1mm}

    def \textit{AAE}($U$, $L$, $\Theta$, $T$): \\
	
	\vspace{1mm}

    \While{U {$\neq \emptyset $} }{
        Sample $S$ from $U$ \\
        $\Theta$ = $\mathcal{P}$ ($S$, $\Theta$) \\
        $U$ = \{$u$ | $f(\Theta, u) < T, u \in U$\}
    }
	return ($\Theta$)
\end{algorithm}

Our active learning method is shown in Algorithm~\ref{algor}. The input of this algorithm consists of four parts, including the unlabeled dataset $U$, the labeled dataset $L$, the parameters of deep classifier $\Theta$ and the threshold of estimation $T$. We repeat the sampling and retraining process until that the unlabeled dataset is empty. In every round, we first randomly sample some data from the unlabeled dataset. And then use these data to train deep classifier. Next, we use the deep classifiers to evaluate the point in $U$. If the evaluation result reaches the threshold, we put this point into $L$. Finally, we return the parameter of the deep classifier. The deep classifier is the key component of our AAE. We consider to construct our classification based estimator by utilizing popular neural networks.

\subsection{Deep classifier}
Nowadays, deep learning express great learning ability in many fields, such as image classification, NLP and etc. The popular deep model mainly contains two basic kinds, convolutional neural networks(CNN) and recurrent neural networks(RNN). And many traditional networks consist of them, such as VGG~\cite{vgg}, RCN~\cite{ref44} and etc. In this paper, considering the different characteristics of different graph dataset, we attempt to employ two kinds of networks to learn the complex graph features introduced in section~\ref{feature}.

\subsubsection{CNN based classifier}
The features of our evaluation task is every complex. For example, there exists significant interaction between the statistical features of graph data and the storage vectors. So, we firstly attempt to use CNN to construct evaluation model.
Compared with NNs(Neural Networks), CNN has more powerful capabilities in learning sectional features. The different convolution kernel filters could learn more mature  features.
And convolutional filters could extract the sectional features of the graph data, graph workload, graph storage solution. Consequently, the multi-layer convolution kernel will learn model detailed knowledge.

\begin{table}[h]
    \caption{The deep classifier based on CNN}
    \begin{center}
    \begin{tabular}{|p{1.1in}|p{2.5in}|p{0.8in}|}
    \hline
    \textbf{Layer Name}&\textbf{Parameter Value}&\\
    \hline
    Input & & ALL \\
    \hline
    Conv1&filters=16,kernel\_size=3&ALL\\
    \hline
    Conv2&filters=16,kernel\_size=3,activation=tanh&ALL\\
    \hline
    MaxPooling1&pool\_size=3&ALL\\
    \hline
    Conv3&filters=16,kernel\_size=3,activation=tanh&ALL\\
    \hline
    Conv4&filters=16,kernel\_size=3,activation=tanh&ALL\\
    \hline
    MaxPooling2&pool\_size=3&ALL\\
    \hline
    Conv5&filters=16,kernel\_size=3,activation=tanh&DCNN\\
    \hline
    Conv6&filters=16,kernel\_size=3,activation=tanh&DCNN\\
    \hline
    MaxPooling3&pool\_size=3&DCNN\\
    \hline
    Flatten&&ALL\\	
    \hline
    Dense&1,activation='sigmoid'&ALL\\
    \hline
    Output & &ALL \\
    \hline
    \end{tabular}
    \label{table5}
    \end{center}
    \end{table}

Considering that the number of layers has great influence on learning ability. In this paper, we design two CNN models, a simpler one and a deeper one, for researching the effect of network structure in graph workload evaluation. As shown in the Table~\ref{table5}, the first simpler CNN model(SCNN) contains four convolutional layers. And the deeper CNN model(DCNN) contains six convolutional layers. For the SCNN, we first use two convolutional layers with 16 filters to extract features in original input vector. And we employ the tanh as the activation function to avoid the gradient disappearance. After two convolutional layers, a maximization pool MaxPooling1D is added to retain the main features. Next, we add two convolutional layers and one pooling layer to enhance the learning ability of our model. Finally, we add the flatten layer to connect the convolutional layer and the fully connected layer. And for the DCNN, we add extra two convolutional layers and pooling layer to improve the capacity of evaluation model.

\subsubsection{RNN based classifier}

 For complex characteristics of graph evaluation task, the previous partial input would influence the later output in current partial input. Therefore, we also consider to utilize RNN to process our evaluation task. Different from traditional NNs, RNNs introduce directional loops, which can deal with the correlation between inputs. The specific manifestation is that the network will memorize the previous information and apply it to the calculation of the current output. However, for traditional RNNs, there are some problems of gradient disappearance and gradient explosion. In order to solve these problems, K et al. proposed GRU\cite{ref19}, a variant of LSTM network\cite{ref20}, is also a type of RNN. And the internal structure of GRU is similar to that of LSTM. So, we utilize the GRU to construct the deep graph estimator.

\begin{figure}[htbp]
    \centering
    \includegraphics[width=3.5in]{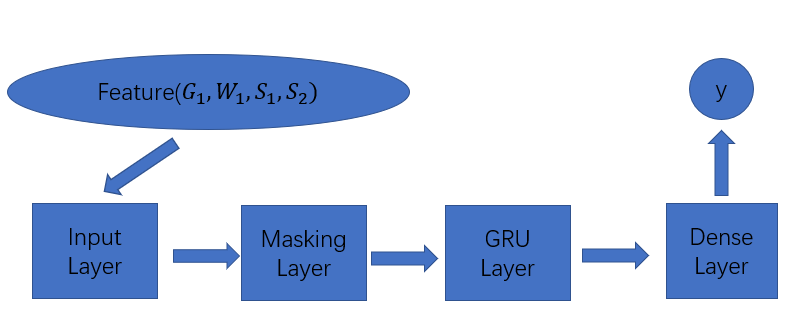}
    \caption{The Deep classifier based on GRU}
\label{figure3}
\end{figure}

As shown in Figure~\ref{figure3}, The GRU classifier consists of four layers. The input layers is responsible for receiving input vector which is support after feature engineering. Because the different graph data, graph workload and graph storage solution have the different features, constructed feature vectors have variable length.  In order to train these vectors in one batch, we implement data preprocessing (fill -1 to vectors) to unify the length of vectors. And correspondingly, we add a mask layer to prevent the influence of the filling operation. Next, we add an GRU layer to learn the evaluation task. Finally, we could obtain the evaluation result after a dense layer.

\section{ Experimental Results}\label{result}
In this section, we express our examination results of our model. First, we introduce the experimental settings in section~\ref{set}, containing dataset and execution environment. Then, we discuss our experimental results from three aspects, active performance in section~\ref{res1}, classifier comparison in section~\ref{res2} and evaluation time cost in section~\ref{res3}. The source codes of our experiments are available~\footnote{https://github.com/yangmanST/Graph-Storage/tree/master}.
\subsection{Experimental Settings}\label{set}
\subsubsection{Dataset}
\begin{table}[h]
    \caption{Features of datasets}
    \begin{center}
    \begin{tabular}{|p{1.6in}|p{1in}|p{1in}|p{0.5in}|}
    \hline
    \textbf{}&\textbf{Freebase-small}&\textbf{Freebase-middle}&\textbf{LDBC}\\
    \hline
    number of nodes&480577&4264156&184328\\
    \hline
    number of edges&314753&3147537&767894\\
    \hline
    number of node types&1&1&8\\
    \hline
    number of edge types&1814&2912&15\\
    \hline
    number of property types&3&3&62\\
    \hline
    \end{tabular}
    \label{table1}
    \end{center}
    \end{table}
In this paper, we utilize two open source graph datasets, LDBC and Freebase in micro-benchmark~\cite{ref14} to conduct our experiments. Micro-benchmark consists of seven synthetic and real dataset. The selected two datasets in our study are the Freebase~\cite{freebase}(a real dataset) and the LDBC dataset~\cite{ldbc}(a synthetic dataset). In Freebase, a subgraph (Frb-O), which only has nodes related to organization, business, government, finance, geography, and military are considered, is created. Then, by randomly selecting 0.1 percent and 1 percent edges of this subgraph, two datasets, Freebase-small and Freebase-middle are derived.
For the LDBC dataset, the synthetic dataset (LDBC) is generated by using the data generator provided by the Linked Data Benchmark Council. The graph it generates simulates the characteristics of a real social network with a power law structure and the characteristics of the real world. Finally, we use different dataset to conduct our experiments, as shown in Table~\ref{table1}.

\subsubsection{Environment}
In this paper, we use two open source graph databases to complete our experiments, as shown in Table~\ref{database}. And our experiments are conducted on the Ubuntu 18.04 with 8G memory and 80G disk. The deep classifier is trained with  CPU and 8G memory.

\begin{table}[h]
    \caption{Features of datasets}
    \begin{center}
    \begin{tabular}{|p{0.6in}|p{0.6in}|p{0.7in}|p{1.2in}|}
    \hline
    \textbf{Name}&\textbf{Version}&\textbf{Storage}&\textbf{Query Language}\\
    \hline
    New4J&1.9 &Native Graph &Gremlin\\
    \hline
    Titan&0.4.5 &Columnar store &Gremlin\\
    \hline

    \end{tabular}
    \label{database}
    \end{center}
    \end{table}

\subsection{Active Performance}\label{res1}
In this section, we introduce the evaluation accuracy with the different rate of training set in different datasets. Table~\ref{activep} reports the classification accuracy of different active stage (58\%, 49\%, 41\%) in Freebase-small, Freebase-middle and LDBC. We train our classifiers with learning rate = 0.01 and traditional BP algorithm. After carefully tuning, our active learned estimator achieves 99.85\% prediction accuracy in LDBC labeled dataset ($L$ in algorithm~\ref{algor}) and reach 79.34\% prediction accuracy in LDBC unlabeled dataset ($U$ in algorithm~\ref{algor}) with only 41\% training dataset. We can see that our active structure could reduce the demand of training set. The estimator still performs well in the low rate of training set. That is, our active method could support efficient evaluation with lower time consumption.

\begin{table}[htbp]
    \caption{The Classification Accuracy in Different TrainSet}
    \begin{center}
    \begin{tabular}{|p{0.6in}|p{0.6in}|p{0.6in}|p{0.5in}|p{0.5in}|p{0.5in}|}
    \hline
    \textbf{Dataset}&\textbf{Train Set Rate}&\textbf{Test Set}&\textbf{SCNN}&\textbf{DCNN}&\textbf{GRU}\\
    \hline
    Freebase-small& 58\% & $L$  &94.19\% &	93.02\% &	91.86\% \\
    \cline{3-6}
    & & $U$ & 68.75\% &	68.75\% &	56.25\% \\

    \hline
    Freebase-middle&49\% & $L$  & 93.25\%& 	82.02\%&85.39\%\\

    \cline{3-6}
    & & $U$ & 72.73\% &	63.64\% & 61.16\% \\

    \hline
    LDBC& 41\% &$L$ &99.85\% &	99.85\% &	90.62\%\\

    \cline{3-6}
    & &$U$ & 78.51\% &	79.34\% & 68.60\% \\

    \hline
    \end{tabular}
    \label{activep}
    \end{center}
    \end{table}

\subsection{CNN \& GRU}\label{res2}
In this paper, we propose three deep classifiers to learning the complex graph workload evaluation task. In this section, we compare the performance of these deep classifiers in two aspects. One is the classification accuracy, and the other is the cross validation in different classifiers.

\begin{figure}[htbp]
    \centering
    \includegraphics[width=4in]{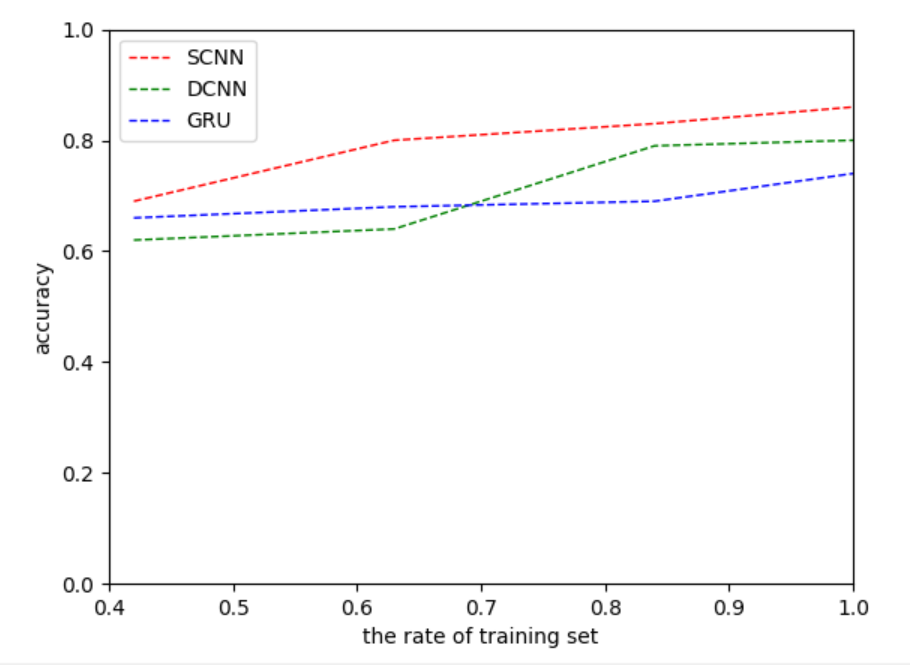}
    \caption{The Accuracy of Unlabeled Dataset (LDBC) on Different classifiers}
	\label{LP}
\end{figure}

\begin{figure}[htbp]
    \centering
    \includegraphics[width=4in]{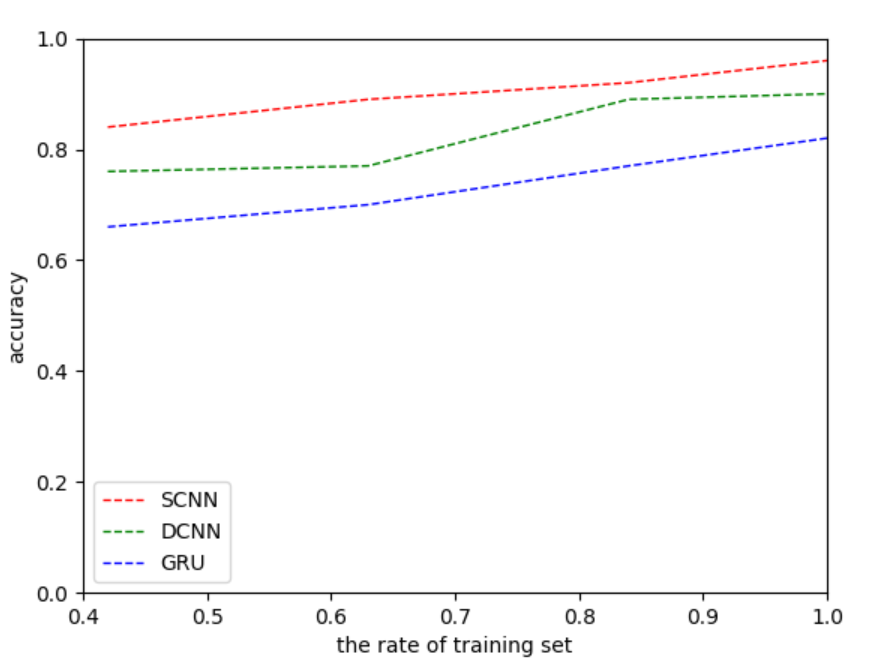}
    \caption{The Accuracy of Full Dataset(LDBC) on Different classifier}
	\label{UP}
\end{figure}

As shown in Figure~\ref{LP}, we test the prediction accuracy in our three classifiers in unlabeled dataset ($U$). 
We can find that SCNN is significantly better than other two classifiers (DCNN and GRU). For SCNN, its accuracy is 10.1\% higher than DCNN and 12.6\% higher than GRU on average. Correspondingly, Figure~\ref{UP} shows the accuracy in three classifiers with full dataset ($L$ + $U$). The accuracy of SCNN is 7.8\% higher than DCNN on average, and 18.3\% higher than GRU.

\begin{figure}[htbp]
    \centering
    \includegraphics[width=4in]{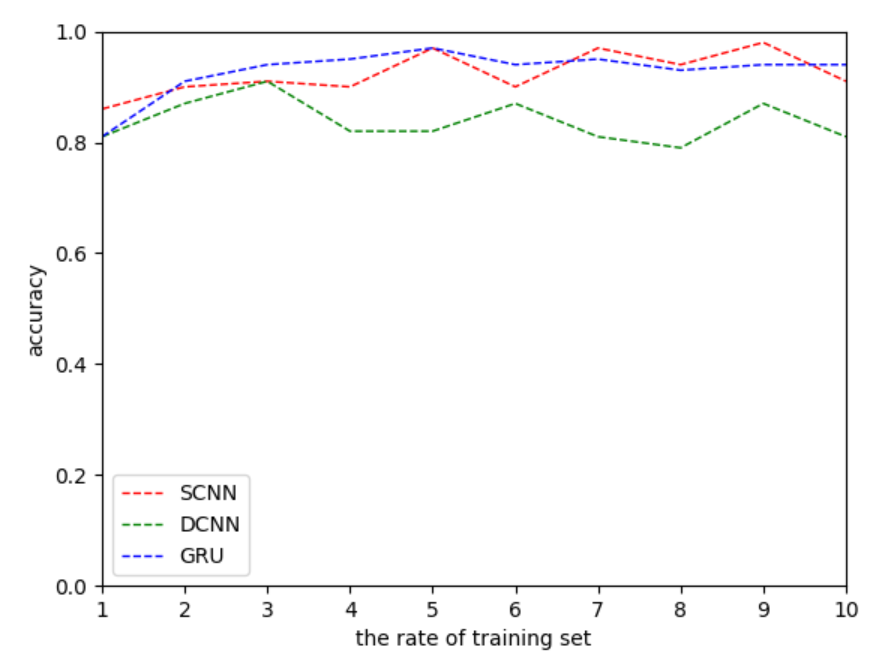}
    \caption{Cross-validation Results of LDBC on Different classifier}
\label{cross}
\end{figure}

Figure~\ref{cross} presents the result of cross-validation of the three classifiers on the LDBC. The average accuracy and its standard deviation of SCNN is 89.8\% and 4.2\%,DCNN is 86.1\% and 4.3\%,GRU is 89.7\% and 4.1\%. It's shown that the performance of SCNN is more stable than other two classifiers.

Comprehensively, compared to DCNN and GRU, SCNN which is expert at extracting sample features and mining the correlation of different features is the best classifier in our graph workload evaluation task. 

\subsection{Evaluation Time Cost}\label{res3}
\begin{figure}[htbp]
    \centering
    \includegraphics[width=4in]{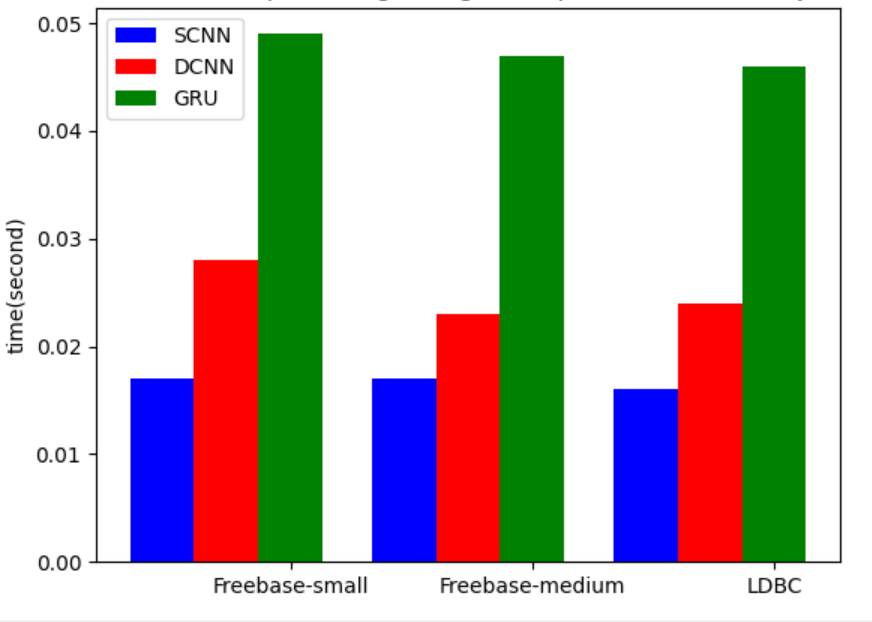}
    \caption{Average Time Cost in Different Dataset(neo4j)}
\label{time1}
\end{figure}

\begin{figure}[htbp]
    \centering
    \includegraphics[width=4in]{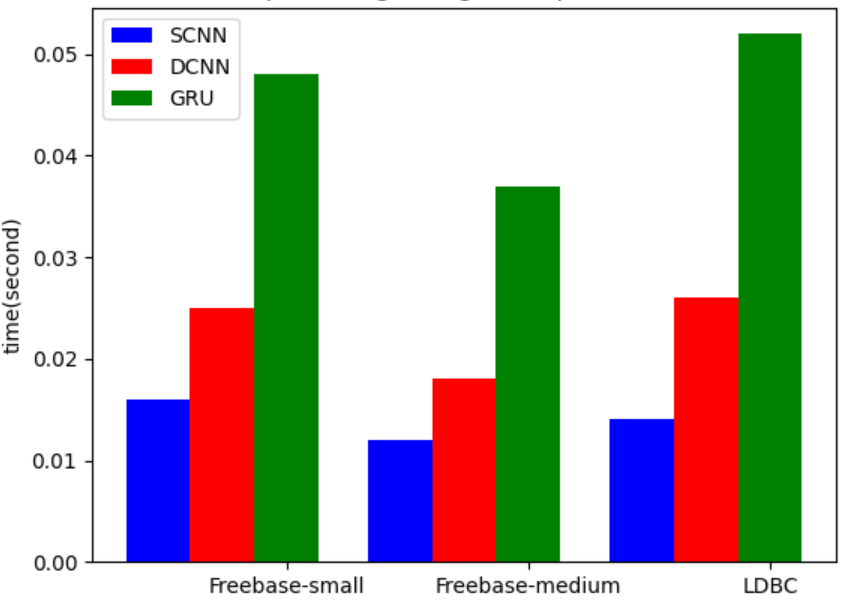}
    \caption{Average Time Cost in Different Dataset(Titan)}
\label{time2}
\end{figure}

In order to process the changing graph workload, the evaluation model should be able to quickly process the evaluation requirements. In this section, we discuss the average evaluation time of our classification-based estimator. As shown in Figure~\ref{time1}, in neo4j, we use 50 data to test the average estimation time in different dataset and classifiers. Freebase-small dataset takes separately 0.017s, 0.017s, 0.016s for graph workload evaluation. Freebase-medium dataset takes separately 0.028s, 0.023s, 0.024s. LDBC dataset takes separately 0.049s, 0.047s, 0.046s. Figure~\ref{time2} shows the average time consumption in Titan. Freebase-small dataset takes separately 0.016s, 0.012s, 0.014s. Freebase-medium dataset takes separately 0.025s, 0.018s, 0.026s. LDBC dataset takes separately 0.048s, 0.037s, 0.052s. We can see that the prediction time is mainly related to the structure of classifiers. Because the runtime of deep learning based model has directly relationship with the number of neurons in classifiers. Our estimator could also quickly predict in any other dataset after some limited training.

\section{Conclusion}
In this paper, we design an active auto-estimator for improving the graph storage solution. We carefully design the feature engineering, including graph data features, graph workload features and graph storage solution. And we clarify the time efficiency and evaluation accuracy in two open source datasets. In the future, we expect to make more improvements in the following two aspects. On the one hand, we want to figure out if the more fine-grained features (such as the graph itself) would performs well than the processed statistical features (such as the number of nodes). After all, it would loss some information after some processing in original dataset.
On the other hand, we would consider to research if the new researches about auto-ML would inspire the graph workload evaluation for obtaining a perfect deep classifier.


\end{document}